\documentclass[12pt]{article}
\usepackage{graphicx}
\usepackage{setspace}
\begin{document}
\begin{center}
CLASSICAL OSCILLATOR WITH POSITION-DEPENDENT MASS IN A COMPLEX
DOMAIN  \vskip 1cm
{Subir Ghosh{\footnote{subir\_ghosh2@rediffmail.com}}}\\
Physics and Applied Mathematics Unit,\\
Indian Statistical Institute, 203 B. T. Road, Kolkata 700108, India\\
\vskip .2cm
 and\\
 \vskip .2cm
 {Sujoy Kumar Modak{\footnote{sujoy@bose.res.in, Ph :91-03323351313/0312, Fax :91-03323359176/3477}}}\\
S.~N.~Bose National Centre for Basic Sciences,\\
JD Block, Sector III, Salt Lake, Kolkata-700098, India.\\
\end{center}
\vskip .2cm
{\bf Abstract:}\\
We study complexified Harmonic Oscillator with a {\it{position-dependent mass}}, termed  as Complex Exotic Oscillator (CEO).  The
complexification induces a gauge invariance
 \cite{sm,mos1}.  The role of $PT$-symmetry is discussed from the perspective of classical trajectories of
CEO for real energy. Some  trajectories of CEO are similar to
those    for the particle in a quartic potential in the complex
domain \cite{Ben2,hook}.\vskip .4cm
PACS: 03.65.Ca, 03.65.Ge

\newpage

{\bf{Introduction:}} It came as a surprise when the early works
\cite{Mc,Graf,Cal,And,Hol,Schol,recent} showed that certain
quantum theories with  complex  Hamiltonian have  real spectra.
Subsequently this observation was explained \cite{Ben,Ben1,Ben2}
from the fact that these systems enjoy the combined $PT$
(parity and time reversal) symmetry. The consistency of these
models as quantum systems was established \cite{mos1} by
constructing a positive definite inner product that generates
unitary evolution. Later on there has been a lot of activity
\cite{Bag1,Most,Bag2,Sin,Zno,Gey,RB} in the study of different aspects of
$PT$-symmetric models.

These models are referred to as ``Crypto''-Hermitian models by Smilga
\cite{sm}. In \cite{sm} Smilga has also provided an alternative
explanation to this behavior (of having real energy eigenvalues
for a complex Hamiltonian) in terms of a gauge invariance.
However, in an important earlier work by Mostafazadeh \cite{ali},
it was observed in a general context that the real part of the
Hamiltonian can generate the dynamics in a real phase space. In addition the imaginary part of the Hamiltonian, treated as a
constraint, can generate symmetry transformations. The usage of
certain class of coordinates in previous works
\cite{Xav,Kau1,Kau2} was also explained in \cite{ali}.

In our analysis we shall follow the method developed in \cite{sm}. The idea used in \cite{sm} is to complexify a real Hamiltonian system and
subsequently treat the real part of the complex ${\cal{H}}$ as the
Hamiltonian $H$ of the enlarged system. As a consequence the number of degrees of freedom of the new real Hamiltonian ($H$) is actually twice 
the original real Hamiltonian. By virtue of Cauchy-Riemann
condition (for ${\cal{H}}$) and Hamiltonian equations of motion it
is possible to show that both the real part $H$ and the imaginary
part $G$ of ${\cal{H}}$ (where ${\cal{H}}=H+iG$), are
{\it{separately conserved}}. Hence $G$ acts as a First Class
Constraint in the terminology of Hamiltonian constraint analysis
of Dirac  \cite{dir} and the presence of $G$ ensures the equality of the degrees of freedom count of the Hamiltonian
 system before and after the complexification. In particular $G=0$ forces the energy to be real. This FCC is present in 
all such complexified systems and the gauge symmetry induced by it \cite{dir} is termed as Crypto-gauge symmetry \cite{sm}.
 In \cite{sm} it has been shown
that specific features of some complexified models, (analyzed in
terms of real variables), can be matched with their $PT$-symmetric
counterpart in the complex plane \cite{Ben2}. The advantage of the
formalism developed in \cite{sm} is that one can start with any
real Hamiltonian model that is convenient and study its complex
generalization.

The study of  Smilga \cite{sm} was generalized to more than  one dimension by one of us in \cite{majhi} for
 the complexified Harmonic Oscillator. It was shown that the straightforward
generalization yielded a richer and qualitatively different constraint structure where both 
the First Class  and Second Class Constraints (SCC)  \cite{dir} are present. The presence of SCC induces a
change in the symplectic structure. The additional constraints, besides $G = 0$, emerge from the demand that
 like energy, the angular momenta should also have real spectra. An interesting
feature was revealed in this study \cite{majhi}: the number of
First Class  and Second Class constraints are such that the
degrees of freedom count before and after complexification remains
unchanged. The present work with a different model - the Complex
Exotic Oscillator (CEO) - (or equivalently complex Harmonic
Oscillator with a {\it{position-dependent mass}}), also behaves in
this way and it is clear that this feature is generic. Recall that
Cauchy-Riemann condition (for ${\cal{H}}$) and Hamiltonian
equations of motion were all that were needed to show that   both
the real part $H$ and the imaginary part $G$ of ${\cal{H}}$ (where
${\cal{H}}=H+iG$), were separately conserved and one can interpret
$G=0$ as a constraint. In particular it was a First Class
Constraint and induced the gauge invariance. However, we still
have not been able to provide an analogous proof for the rest of the
constraint algebra in higher dimensions.

Exotic Oscillator (EO) - the parent model of the present study -
has an interesting history. Similar models have been studied
before in the guise of a Harmonic Oscillator with a position
dependent mass \cite{mat, Cari1,Cari2,Cari3, Proy1,Tez}. The model
has also been quantized \cite{Cari1,Cari2,Cari3}. Our present work
deals with the classical setup. We complexify the EO to
obtain the CEO and show that in higher (three) dimensions the
constraint structure is same as that of the Complex Harmonic
Oscillator \cite{majhi}. The major part of the paper deals with
the CEO obtained from the one dimensional EO where we plot the
trajectories of the path for fixed energy. It is found that the trajectories of the CEO are nontrivial
 generalization of the trajectories of the Complex Harmonic Oscillator \cite{sm} and one can interpret the latter as a
 special case of the former. Incidentally in some cases we find a close
similarity between the trajectories presented here with those of
the complex anharmonic (quartic) potential \cite{hook,Ben,Ben1}
{\footnote{We thank Professor Hook for pointing this to us.}}.

We  mention the role of $PT$-symmetry in our work and 
later provide a discussion on the possible classical analogue of  Exceptional
points  \cite{kato, heiss, dorey, znojil} in the present study. We
follow the definition provided in \cite{hook} for $PT$-symmetry in
classical mechanics in the context of our Hamiltonian  system that yields a set of ordinary differential equations. Under $PT$-transformation,  a generic complex function $f(t)$ is transformed to  $f(-t)^*$ such that  for $PT$-symmetric functions $f(t)=f(-t)^*$. Furthermore, an ordinary  differential equation $\ddot f(t)=F(\dot f(t),f(t))$  will be  $PT$-symmetric if for every solution $f(t)$ of this differential equation, $f(-t)^*$ is also a solution of the same equation. The same definition will hold for a system of ordinary differential equations.  In general the solution of a differential equation may not respect the symmetry the differential equation enjoys. In case of $PT$-symmetric 
systems the solution will also be $PT$-symmetric if it remains
invariant under reflection with respect to the imaginary axis. Now
in the formalism used here \cite{sm} the variables we deal with
are real and the complex conjugation of the solution does not play
any role. However its role is taken up by the extra variables that
appear in the process of complexification and $PT$-symmetry nature
of a solution is judged in a similar way by noting if it is
symmetric under reflection with respect to the axis related to the
extra variable that comes from the complex sector. This will
become clear as we proceed to draw the solutions or particle
trajectories later in the paper.

Although quantization of the  classical model studied here remains
outside the scope of our paper we would like to mention the
following point: for quantization of a constrained system there is
some amount of non-uniqueness involved since the two acts,
{\it{i.e.}} imposition of the constraints and quantization are not
commutative (see \cite{sm} for more details).

{\it{\bf The Exotic Oscillator and its Complexification}}: We start by describing the Exotic Oscillator (EO). The
results presented here are not new and can be extracted
from the more general settings provided in
\cite{mat,Cari1,Cari2,Cari3}. We only reproduce it in the simplest
model to highlight the contrasting behaviors of Harmonic and
Exotic Oscillators. Consider the Hamiltonian, $
H=\frac{1}{2}(p^2-b(xp)^2) $ where $(xp)=x_i p_i,~x^2=x_ix_i,~
p^2=p_ip_i; ~i=1,..,n$ and the particle mass is taken to be
unity. It has the interesting dynamics $ \ddot x_i=-2b
(\frac{1}{2}(p^2-b(xp)^2))x_i=-(2bH)x_i $ where the Hamiltonian ($H$) itself 
appears in the place of the frequency parameter for a normal Harmonic Oscillator {\footnote{For the sake of
curiosity, we note that in general, for
$$
H=[\frac{1}{2}(p^2-b(xp)^2)]^\nu \equiv h^\nu  ~~\Rightarrow \ddot
x_i=-(2\nu^2bh^{(2\nu -1)})x_i $$ for a constant $\nu $ .}}.

 The above system is classically symmetric under the $PT$-transformations as
defined in Introduction. The  Hamiltonian  can be obtained from
the Lagrangian $ L=\frac{1}{2}\dot x^2 +\frac{b(x\dot
x)^2}{2(1-bx^2)} $. A particular solution $ x_i=A_iSin(\omega t)
$ obeying the  condition $ bA^2=1$ leads to the time independent Hamiltonian $
H=\frac{1}{2}\omega ^2A^2=\frac{\omega ^2}{2b}.$

Hence the above solution  is bounded with the parameter space of
$A_i$ being a (hyper-) sphere $ A^2=\frac{1}{b} $ but notice that
the phase space motion of EO is unbounded since $ p_i=\frac{\omega
}{Cos (\omega t)}A_i. $ On the other hand a  normal
 Harmonic Oscillator (HO) $
H=\frac{1}{2}(p^2+ax^2)$ has  a solution $ x_i=A_iSin(\omega t) $
for $ \omega ^2=a $ with $H=\frac{1}{2}\omega ^2A^2$.  Hence for
HO there is no restriction on the amplitude $A_i$ but the frequency
$\omega $ depends on the parameter `$a$' present in the HO
Hamiltonian. This is in contrast to the EO case (as shown above)
where there is no restriction on  frequency $\omega $ but the
amplitudes $A_i$ are not arbitrary. It depends on the parameter
`$b$' present in the EO Hamiltonian. Also, the phase space
trajectories are bounded and unbounded for HO and EO respectively.

In the present work we shall consider some aspects of phase space
profile when we complexify the EO. This needs a stabilizing term
for the EO and the simplest remedy is to introduce the Harmonic
Oscillator potential with strength `$a$', such that
\begin{equation}
H=\frac{1}{2}(p^2-b(xp)^2 +a x^2), \label{14}
\end{equation}
with $ \ddot x_i =-[2bH + a(1-2bx^2)]x_i. $ Now we can compare the
relative effects of the two interaction terms simply by tuning the
parameters `$a$' and `$b$'.  Clearly this extended system also
enjoys  $PT$-symmetry. In the following analysis we deal with the
Complex extension of EO and classify the constraints particularly for three dimensional case.

 We  consider the CEO  in the prescribed way \cite{sm,majhi} by replacing $x_i,p_i$ by $z_i,\pi _i$ respectively, in the
 Hamiltonian (\ref{14}), to yield
 \begin{equation}
{\cal{H}}(\pi_i,z_i)=\frac{1}{2}(\pi^2-b(z\pi)^2 +a z^2).
\label{14a}
\end{equation}
 The identification of $z_i,\pi _i$ is given in terms of real canonical
degrees of freedom as $ z_i = x_i+iy_i~;~~ \pi_i=p_i-iq_i$ and they satisfy the relations $\{x_i,p_j\}=\delta_{ij}~~;~~ \{y_i,q_j\}=\delta_{ij}$. The
complex Hamiltonian  ${\cal {H}}$ in (\ref{14}) now reads,
\begin{eqnarray}
{\cal{H}}(\pi_i,z_i) = H(p_i,q_i;x_i,y_i)+iG(p_i,q_i;x_i,y_i)
\label{1.3}
\end{eqnarray}
where the real and imaginary parts are respectively,
\begin{eqnarray}
H={\frac{1}{2}}{\left[(p^2-q^2)+a(x^2-y^2)-b[(xp)^2 + (yq)^2 -
(xq)^2 - (yp)^2
 + 2(xp)(yq) + 2(xq)(yp)]\right]},
\end{eqnarray}
\begin{eqnarray}
G=-(pq) + a(xy) - b[(xp)(yp) - (xp)(xq) + (yp)(yq) - (xq)(yq)].
\end{eqnarray}
Customarily the real part $H$ is taken as the Hamiltonian that
generates the time evolution. The real and imaginary part satisfy the relation $\{G,H\}=0$.  We further restrict the system by
imposing the constraint,
\begin{eqnarray}
G \approx 0, \label{01.5}
\end{eqnarray}
(The weak equality is interpreted in the
sense of Dirac \cite{dir}.) This simplifying choice is same as
that of \cite{sm} but indeed one can (and should) consider more
general models with $G$ being some non-zero $c$-number.

 In a  previous work by one of us \cite{majhi} we considered the complex Harmonic oscillator and discussed the full constraint algebra by including the angular momentum as well. The angular momentum, given by $L_i=\epsilon_{ijk}x_jp_k$, satisfy the relation $~~\dot L_i=\{L_i,H\}=0$ in both the cases. Interestingly the rest of the algebra is also repeated here although the form of $H$ and $G$ are different.

Lastly, it is worthwhile to  study the equation of motion:
\begin{eqnarray}
\ddot{x_i}=-[a+2bH-2abx^2+2aby^2]x_i + 2b[G-2a(xy)]y_i
\end{eqnarray}
Notice that for $a=0$, that is without the Harmonic Oscillator
potential term we obtain,
$$\ddot{x_i}=-[2bH]x_i + 2b[G]y_i.$$ This means that even for the CEO, the characteristic feature of the dynamics
is preserved on the constraint surface where $G=0$. 

{\bf Dynamics of one-dimensional Complex Exotic Oscillator (CEO):} We study the CEO (\ref{14}) in one dimension,
\begin{eqnarray}
H=\frac{1}{2}p ^2 (1-bx^2)+ \frac{a}{2}x^2.
 \label{4}
\end{eqnarray} Obviously the system is considerably simplified. Also
it is clear to see that it constitutes an oscillator with a
position dependent mass. The equation of motion is $ \ddot
x=-2[bH-a(bx^2-\frac{1}{2})]x $. The $PT$-symmetry of the Hamiltonian is manifest.

We now extend the system to complex domain with $z = x+iy~;
\pi=p-iq$,
\begin{eqnarray}
{\cal{H}}=\frac{1}{2}\pi ^2 (1-bz^2)+ \frac{a}{2}z^2.
 \label{1a}
\end{eqnarray}
The real and imaginary parts of ${\cal{H}}$ respectively are
\begin{eqnarray}
H=\frac{1}{2}[a(x^2-y^2)+(p^2-q^2)(1-b(x^2-y^2))-4bxypq],
 \label{ch}
\end{eqnarray}
\begin{eqnarray}
G=xy(a-b(p^2-q^2)) -pq(1-b(x^2-y^2)).
 \label{c}
\end{eqnarray}
In one dimension $G$ is the only constraint since $\{G,H\}=0$ and
it is a First Class Constraint inducing gauge invariance. We consider the case where $G=0$.
 The Hamiltonian equations of motion are,
\begin{equation}
\dot x=p(1-b(x^2-y^2))-2bxyq,
\label{new1}
\end{equation}
\begin{equation}
\dot p=bx(p^2-q^2)+2bypq -ax,
\label{new2}
\end{equation}
\begin{equation}
\dot y=-q(1-b(x^2-y^2))-2bxyp,
\label{new3}
\end{equation}
\begin{equation}
\dot q=-by(p^2-q^2)+2bxpq+ay .
 \label{new4}
\end{equation}
For constant energy $H=E$ the variables $(x,p,y,q)$, must satisfy
the conditions,
\begin{equation}
E=\frac{1}{2}[a(x^2-y^2)+(p^2-q^2)(1-b(x^2-y^2))-4bxypq],
\label{eo1} \end{equation}
\begin{equation}
 xy(a-b(p^2-q^2)) -pq(1-b(x^2-y^2))
=0. \label{eo2} \end{equation} These will help us in determining
consistent  initial conditions when we compute and sketch the
trajectories (albeit numerically). \vskip .7cm
{\bf Trajectories of one-dimensional Complex Exotic Oscillator:} 
We now discuss some features of the trajectories. In the energy
expression   (\ref{eo1}) there are three parameters $E,~a$ and
$b$. Another free parameter `$c$' will appear from our choice of
initial conditions. Throughout our analysis we will keep $\mid E
\mid =\frac{1}{2}$ (since both positive and negative values of
energy can be considered) and put $a=1$ (that is strength of the
stabilizing harmonic potential is fixed to unity). This will allow
us to vary `$b$' (the strength of the exotic term $(xp)^2$) and
`$c$', the parameter that comes from fixing the initial
conditions, mentioned above. Essentially `$c$' signifies the
freedom that we have in the complex domain so that we can choose
initial conditions that allow the CEO to traverse paths which is forbidden for the EO,
 (the latter being restricted to the real space). 
This is the origin of the nested set of trajectories
where in the core we have the real system confined to the real
line in $PT$-symmetry framework \cite{Ben,Ben1} (or real variable
in the present case \cite{sm}). To be more specific, changing
`$c$' is connected to a gauge transformation in the language of
Smilga \cite{sm} that goes from one trajectory to another in the
nested set of trajectories with the same physical parameters. We find that the trajectories are very sensitive to the initial conditions ({\it{i.e.}} the value of $c$). This feature has been stressed earlier in \cite{milt}.

 In Figure 1.1 and Figure 1.2, we study the phase space trajectories ($x$ in abscissa and $p$ in ordinate) of the
 {\it{real}} exotic oscillator (\ref{4}), $$ H=\frac{1}{2}p ^2 (1-bx^2)+ \frac{a}{2}x^2 \equiv 1=p ^2 (1-bx^2)+ x^2$$
 for positive and negative values of $b$ respectively. For our choice of parameters, `$b$' has an upper bound $+1$ for
 positive values. This is because for $b=1$ the solution depicts isolated points $p=\pm 1, x=\pm 1$, and for larger
 positive `$b$' there are no closed orbits. On the other hand,  there is no restriction on `$b$' for negative values.
 For both Figure1.1 and Figure 1.2, $b=0$ reproduce normal ellipse which we expect for the Harmonic Oscillator.
 In Figure 1.1 the limit of `$b$' is $0\leq b<1$ and we find that the trajectories are diverging outside the normal ellipse,
  whereas in Figure 1.2, `$b$' goes up to $b=-1500$, and trajectories are converging inside the normal ellipse.
  Therefore the presence of the parameter `$b$' in the Hamiltonian is realized through the distortion in $(x,p)$
  phase diagram of the real exotic oscillator.

Now we come to rest of the figures where the CEO is studied. From
comparing the works of \cite{sm} (that deals with
 complexified models) and \cite{Ben2} (that studies $PT$-symmetric models), it is clear that in our case the nature of the
 trajectories, as far as $PT$-symmetry is concerned, can be ascertained from the geometrical symmetry of the profiles.
 In all
 the figures we plot $x$ in abscissa and $y$ in ordinate following  our convention $z=x+iy$. Hence, similar to \cite{Ben2}
  where real and complex parts of the coordinate were plotted in abscissa and ordinate respectively, {\it{ trajectories
   that are $PT$-symmetric will be invariant under reflection about the ordinate}}. Note that this is same as the
 trajectories studied in \cite{sm} as well.

In Figure 2.1 we first reproduce the simple nested ellipses for
the complexified Harmonic oscillator with $b=0$, $E=+\frac{1}{2}$
\cite{sm,Ben2}. In Figure 2.2 similar example for CEO for positive non-zero $b$ and $E=+\frac{1}{2}$ are studied. It is clear that the non-zero exotic parameter `$b$' distorts the concentric ellipses.
Trajectories for  negative $b$ are depicted in Figure 2.3. We notice a close similarity between Figure 2.3 and the anharmonic oscillator $(H=\frac{1}{2} p^2+x^4)$ in the complex plane  studied in \cite{hook}. The graphs for negative energy is simply obtained by rotating the figures by $\frac{\pi}{2}$. In Figure 2.4 we fix the initial condition for $E=+\frac{1}{2}$ to get one particular trajectory for given parameter values. The generic form of the initial condition (as referred by {\it initial condition $(A)$} in Figure 2) for these trajectories for positive and negative energies are $$x=c,~p=0,~y=0,~q=\sqrt{\frac{c^2-1}{1-bc^2}},$$ $$x=c,~p=0,~y=0,~q=\sqrt{\frac{c^2+1}{1-bc^2}},$$ respectively,
where $c$ is the free parameter we choose. All the trajectories with such initial condition are $PT$-symmetric closed orbits. It is interesting to note that the trajectory for $b=-0.2499$ in Figure 2.4 is identical to the limiting double cardioid depicted in \cite{Ben2} for $H=p^2+x^2(ix)^\epsilon,~\epsilon =2$.

In Figure 3 we choose a structurally different initial condition (as referred by {\it initial condition $(B)$} in Figure 3 and 4) of the form $$x=1,~p=c,~y=1,~q=\sqrt{c^2+\frac{1+4b}{1+4b^2}},$$
$$x=1,~p=c,~y=1,~q=\sqrt{\frac{b-1}{c(4b^2-1)}},$$ for positive
and negative energies respectively. We notice that the orbits are
not $PT$-symmetric and in general are not closed. However,
for some specific choice of parameters (as in Figures 3.2) we do find closed orbits that are not $PT$-symmetric. 
This is interesting because as mentioned in \cite{hook}, (where examples of this type of orbits are shown for Harmonic Oscillator on complex domain), this occurrence is quite rare. On the other hand, Figure 3.3 is an example of an open orbit which is {\it PT-} symmetric in a restricted sense. Taken as a snapshot over several ``rotations'' one finds an overall left-right reflection symmetry but since the orbit spirals inwards and outwards strict $PT$-symmetry is not maintained. Hence we feel that further studies are needed before one can make a general connection between closeness (openness) of trajectories with presence (absence) of $PT$-symmetry.

However, in the context of $PT$-symmetric quantum mechanics this
type of broken $(PT)$ symmetry situation is connected to the
presence of Exceptional Points \cite{kato,heiss} where a singularity
occurs in the parameter space due to the coalescence of two energy
levels along with their wave functions. Note that this is distinct
from a normal degeneracy where the wave functions of the
degenerate levels are different. Characteristic features of the Exceptional Points survive in the classical trajectories of the corresponding classical problem \cite{heiss,dorey,znojil,milt,smilga}. In fact Exceptional Points can be studied and observed in purely classical systems \cite{dheiss}. (Indeed it is all the more interesting if there exists a corresponding solvable quantum system.) In a classical model of
two coupled damped oscillators this has been discussed by  Heiss
 \cite{heiss}. One can try to put our work in
this perspective by interpreting our dynamical system (\ref{new1} to \ref{new4})
as a very complicated non-linear extension of the two oscillator
model of Heiss \cite{heiss}. We provide an example of an Exceptional Point in Figures 4.1 and 4.2 where the values of $b$ are different. Note that the nature and structure of the two trajectories differ drastically although the relevant points in parameter space are extremely close to each other. We interpret the point $b=-0.18,c=0.8$ (in {\it initial condition $B$}) as an Exceptional Point since at this point the trajectories change from being strictly $PT$-symmetric (Figure 4.1) to non $PT$-symmetric (Figure 4.2). A difference between our system and the example studied by Heiss \cite{heiss} is that the energy in our system does not become complex at any time since the choice of constraint $G=0$ is always present. As we have mentioned before, a more general choice for $G$ may lead to interesting consequences. It would be worthwhile to pursue this problem
further. \vskip .7cm

{\bf Discussions:} The discovery that one can replace the requirement of
hermiticity to that of $PT$-symmetry in order to construct a
Hamiltonian that supports unitary time evolution of quantum states
and real energy eigenvalues has opened the possibility for
considering various types of Hamiltonians that were rejected
before on the grounds of non-hermiticity. Indeed, existence of
$PT$-symmetry plays a role not only in models that are explicitly
non-hermitian and violates either $P$ or $T$ symmetry
individually, $PT$-symmetry is also relevant for models with
Hamiltonians that are individually $P$ or $T$ symmetric. These models, such
as Harmonic Oscillator or particle in a quartic potential exhibits
closed trajectories that are $PT$-symmetric. It is rare that one
finds \cite{hook} trajectories that are closed but not
$PT$-symmetric. The complexified Exotic Oscillator that we have
studied in the present article falls in this category.

We have studied the complexified  Harmonic Oscillator with a
specific form of position-dependent mass. We have referred to it
as Complex Exotic Oscillator. Our motivation for studying this
particular form of effective-mass oscillator (in real space) is
due to the fact that it is a $PT$-symmetric model (in the
classical sense).  It has an interesting form of Hamiltonian
dynamics and furthermore is exactly solvable as a quantum system.

 We find that in general it is possible to have both open and closed
 trajectories for particles. The closed trajectories are mostly
 $PT$-symmetric but we have also found the example of closed orbits that are not $PT$-symmetric. We point out some possible connection of this phenomenon with classical analogue of Exceptional points, that are associated with quantum mechanics when there is a crossover of eigenvalues from real to complex domain.
Some of the {\it PT-} symmetric  orbits resemble closely the orbits of a particle in a quartic potential
 extended in the complex domain \cite{hook}.

Our aim is to study the quantized version \cite{Cari1,Cari2,Cari3} of the classical Exotic Oscillator in the context of Exceptional Point in order to see if the present findings can be corelated.

\vskip.5cm {\bf Acknowledgments}: We are grateful to the referees
for their critical but constructive comments. One of the authors (SKM) thanks
the Council of Scientific and Industrial Research (CSIR),
Government of India, for financial support.

\newpage
\bf{Collected Figure Captions:}

Figure 1: {\it{x vs. p plot:  Distortion of the (x,p) phase
diagram for positive and negative values of b and fixed
$E=\frac{1}{2}$. In Fig $(1.1)$ the limit of $b$ is $0\leq b<1$ and in Fig $(1.2)$ b is varying up to $b=-1500.$}}

Figure 2: {\it{x vs. y plot: Family of trajectories for fixed $E=\frac{1}{2}$ and subjected to fixed initial condition $(A)$; In Fig $(2.1)$ value of $b$ is zero; In Fig $(2.2)$ $b=0.8$; Fig $(2.3)$ represents the family of trajectories for negative $b$ $(b =-10)$; Fig $(2.4)$ is one particular trajectory for $(A)$ and parameter values $c = 2$, $b =- 0.2499$.} }
 
Figure 3: {\it{x vs. y plot: Family of open (Fig $3.1$) and closed (Fig $3.2$) orbits lacking PT symmetry for fixed $E=\frac{1}{2}$ and satisfying the initial condition ($B$). In Fig $(3.1)$ $c=0.06,b=0.5$; In Fig $(3.2)$ $c=0.06, b=1$; In Fig $(3.3)$ $b=1, c=1.2$.}}
 
Figure 4: {{\it{x vs. y plot: Family of trajectories showing the Exceptional Point where the nature of the trajectory changes abruptly; In Fig $(4.1)$ $b=-0.09998, c=0.8$; In Fig $(4.2)$  $b=-0.18, c=0.8$. They both satisfy the fixed initial condition ($B$).}}

\newpage
\begin{figure}[h]
\centering
\includegraphics[angle=0,width=12cm,keepaspectratio]{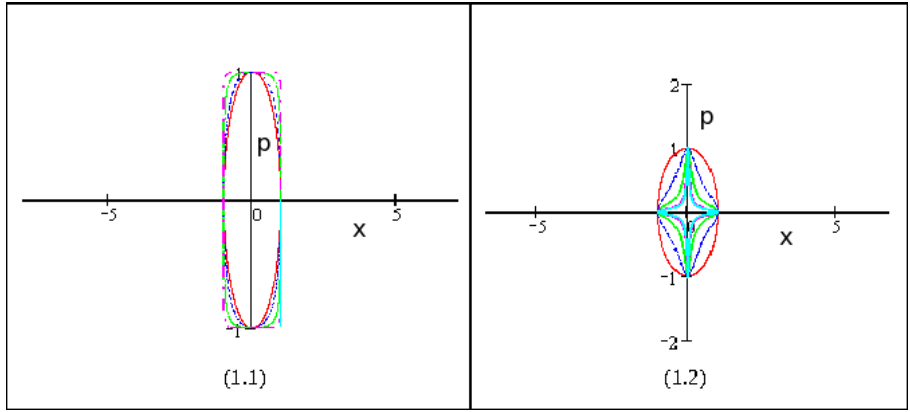}
\caption[]{\it{x vs. p plot: Distortion of the (x,p) phase diagram
for positive and negative values of b and fixed $E=\frac{1}{2}$.
In Fig $(1.1)$ the limit of $b$ is $0\leq b<1$ and in Fig $(1.2)$ b is
varying up to $b=-1500.$}}
\label{fig1}
\end{figure}

\newpage
\begin{figure}[h]
\centering
\includegraphics[angle=0,width=12cm,keepaspectratio]{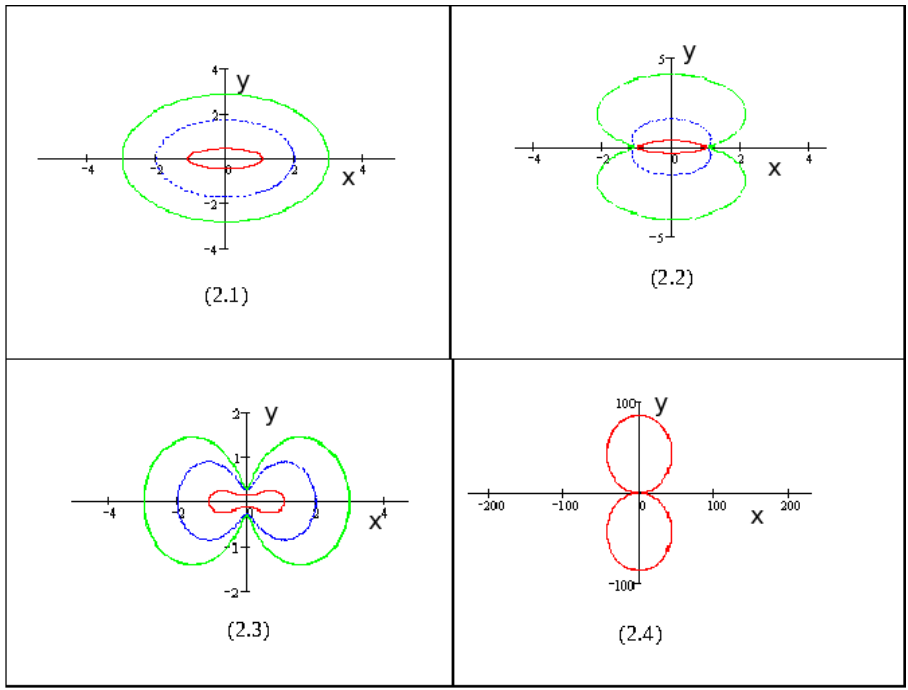}
\caption[]{\it{x vs. y plot: Family of trajectories for fixed $E=\frac{1}{2}$ and subjected to fixed initial condition $(A)$; In Fig $(2.1)$  value of $b$ is zero; In Fig $(2.2)$ $b=0.8$; Fig $(2.3)$ represents the family of trajectories for negative $b$ $(b =-10)$; Fig $(2.4)$ is one particular trajectory for parameter values $c = 2$, $b =- 0.2499$.} }
\label{fig2}
\end{figure}

\newpage
\begin{figure}[h]
\centering
\includegraphics[angle=0,width=10cm,keepaspectratio]{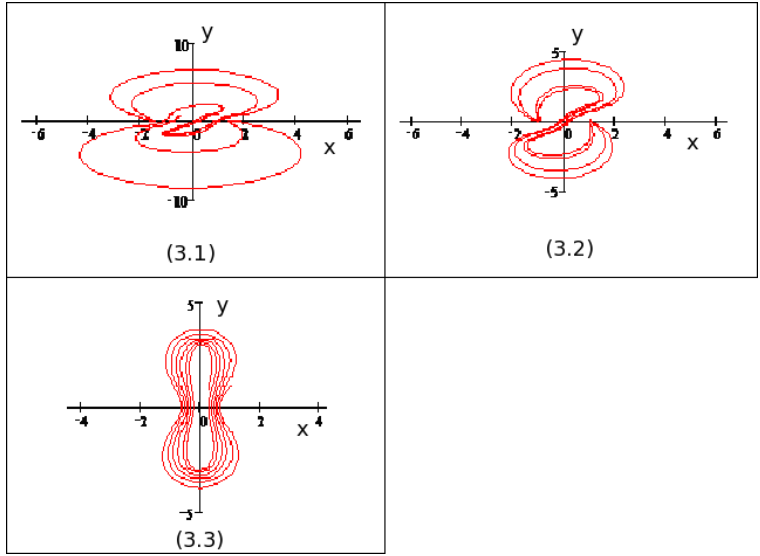}
\caption[]{\it{x vs. y plot: Family of open (Fig $3.1$) and closed (Fig $3.2$) orbits lacking PT symmetry for fixed $E=\frac{1}{2}$ and satisfying the initial condition ($B$). In Fig $(3.1)$ $c=0.06,b=0.5$; In Fig $(3.2)$ $c=0.06, b=1$; In Fig $(3.3)$ $b=1, c=1.2$.}}
\label{fig3}
\end{figure}

\newpage
\begin{figure}[h]
\centering
\includegraphics[angle=0,width=10cm,keepaspectratio]{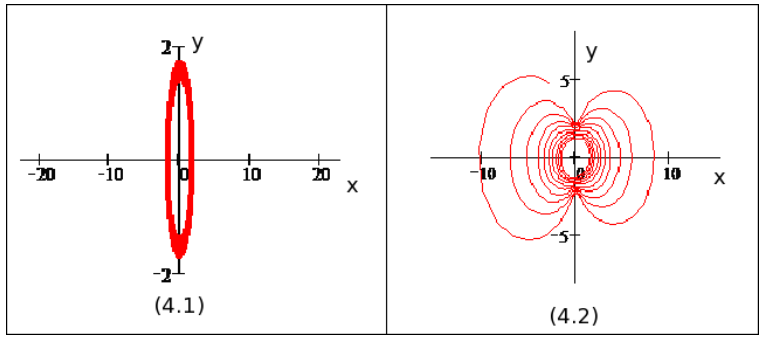}
\caption[]{{\it{x vs. y plot: Family of trajectories showing the Exceptional Point where the nature of the trajectory changes abruptly; In Fig $(4.1)$ $b=-0.09998, c=0.8$; In Fig $(4.2)$  $b=-0.18, c=0.8$. They both satisfy the fixed initial condition ($B$).}}
}
\label{fig4}
\end{figure}

\end{document}